\documentstyle[aps,prl,multicol]{revtex}
\begin{document}
\draft
\title{Low Energy Effective Hamiltonian for the
Fractional Quantum Hall Effect}
\author{MYUNG-HOON CHUNG${}^{a}$, SO-NAM CHOI${}^{b}$,
and JAE-HOON KWON${}^{b}$}
\address{${}^{a}$College of Science and Technology, Hong-Ik University,
Chochiwon, Choongnam 339-800, Korea}
\address{${}^{b}$Department of Mathematics,
Seoul National University, Seoul 151-742, Korea}
\date{\today}
\maketitle
\widetext
\begin{abstract}
A low energy effective Hamiltonian for the fractional quantum Hall
effect is obtained by using irreducible representations of the
symmetry group. It is found that
the model described by the effective Hamiltonian is
similar to the Heisenberg spin chain.
Symmetries of the effective Hamiltonian are studied in order to
decompose the relevant Hilbert space.
This decomposition will be useful for further numerical
analysis on the gap of the quantum Hall system.
\end{abstract}


\raggedcolumns

\begin{multicols}{2}


Strongly correlated systems in low dimensions are of interest recently.
Examples of such systems are noted as spin chain, spin ladder,
the Hubbard model, and so on. In particular,
strongly correlated two-dimensional electron systems
display fascinating phenomena named as the fractional quantum Hall effect (FQHE).

In order to understand the FQHE, a second quantized Hamiltonian was introduced by
Yoshioka, Halperin and Lee\cite{1}.
Diagonalizing the Hamiltonian numerically,
they found a downward cusp in the ground state energy at $\nu=\frac{1}{3}$.
This indicates that the Hamiltonian can be a candidate for the starting point
to explain the physics of the FQHE.
Another important contribution to understanding the QHE
was made by the non-perturbative approach
using the topological invariant\cite{2}.
In this topological approach on a torus geometry,
it is argued that the many-particle ground
state for the FQHE should have the proper degeneracy.
This degeneracy was analyzed by Haldane\cite{3},
who used the many-particle translational symmetries.

As one of strongly correlated systems,
a one-dimensional spin chain shows a peculiar property of ground state\cite{4}.
It is known that the so called Haldane gap is observed in an integer spin chain,
while the gap is absent in a half integer spin chain.

In this paper, using symmetries and irreducible representations,
we derive a low energy effective Hamiltonian
from the Yoshioka-Halperin-Lee Hamiltonian.
During the procedure, the perturbation theory for a degenerate case is adopted.
It is found that the effective Hamiltonian obtained here corresponds to the
model, which is similar to spin chain.
For further numerical calculations, symmetries of the effective
Hamiltonian are discussed.


We begin by considering the second quantized many-particle Hamiltonian,
which is written as
\[
H_{\mbox{red}}=\sum_{a,b=0}^{N-1}W(a,b)\sum_{k=0}^{N-1}c_{k+a}^{\dagger}
c_{k+b}^{\dagger}c_{k+a+b}c_{k},
\]
where $c_{k+N}^{\dagger}=c_{k}^{\dagger}$.
The degeneracy of the Landau level is denoted by $N$.
We call it as the Yoshioka-Halperin-Lee Hamiltonian\cite{1,5}. Here,
we drop out the trivial constant energy term,
$\frac{1}{2}\hbar \omega \sum_{i} c_{i}^{\dagger}c_{i}$,
in the Hamiltonian. The couplings $W(a,b)$ are written as
\begin{eqnarray*}
W(a,b)=\frac{1}{4}\{V(a,b)-V(b,a)\},~~~~~~~~~~~~~~~~ \\
V(a,b)=\frac{e^{2}}{L}\sum_{l,j\in Z}
\frac{\exp[-\frac{\pi}{N}\{(Nl+a)^{2}+j^{2}\}]}{\sqrt{(Nl+a)^{2}+j^{2}}}
\cos(\frac{2\pi}{N}jb),
\end{eqnarray*}
where the case of zero denominator is excluded\cite{5}.

For a QHE system of $M$ electrons, a typical basis of the Hilbert space
${\cal H}(N,M)$ is expressed by creation operators:
$\{c^{\dagger}_{i_{1}}
\cdots c^{\dagger}_{i_{M}}|0>~|~0\leq i_{1} < \cdots < i_{M} \leq N-1 \}$.
The dimension of ${\cal H}(N,M)$ is
given by ${}_{N}C_{M}$.

The essential manipulation to find an effective Hamiltonian
is following.
Let us consider the case of $\nu=M/N=pGD/qGD$,
where $G$ and $D$ are big integers.
It is important to notice that we can rewrite the Hamiltonian by
dividing the system into $D$ pieces. Let
\[
a^{\prime}=Da+A,~~
b^{\prime}=Db+B,~~
k^{\prime}=Dk+K,
\]
where $0 \leq ~a^{\prime},b^{\prime},k^{\prime}~ \leq N-1$,
$0 \leq ~a, b, k~ \leq qG-1$,
and $0 \leq ~A, B, K~\leq D-1$.
Then the original Hamiltonian is now divided as
\[
H_{\mbox{red}}=\sum_{K=0}^{D-1}H_{K00}
+2\sum_{K=0}^{D-1}\sum_{A=1}^{D-1}
H_{KA0}+\sum_{K=0}^{D-1}\sum_{A,B=1}^{D-1}
H_{KAB},
\]
where the part characterized by three indices is written as
\begin{eqnarray*}
&&~~~~H_{KAB}
=\sum_{a,b=0}^{qG-1}
W(Da+A,Db+B)
\sum_{k=0}^{qG-1} \\
&&\times c_{D(k+a)+K+A}^{\dagger}
c_{D(k+b)+K+B}^{\dagger}
c_{D(k+a+b)+K+A+B}
c_{Dk+K}.
\end{eqnarray*}
We notice that each Hamiltonian $H_{K00}$ has the similar
structure of the original $H_{\mbox{red}}$.
In fact, there is a concept of scaling in this game.
Bearing the scaling in mind, we take the following strategy
to find the ground state and the energy gap of the
Hamiltonian $H_{\mbox{red}}$.
The strategy is that we first solve $\sum H_{K00}$
exactly, and then use the perturbation theory for a degenerate case
in order to treat other parts.

We have divided the Hamiltonian into three parts: the first one
playing like a free Hamiltonian, and the other two playing as interactions.
The corresponding Hilbert space is now replaced as
\[
{\cal H}(N,M) \rightarrow \bigoplus_{M_{0}+\cdots+M_{D-1}=M}
(\bigotimes_{K=0}^{D-1}{\cal H}(K;\frac{N}{D},M_{K})),
\]
where $M_{0},\cdots,M_{D-1}$ represents the numbers of electrons in
the $D$ pieces, respectively.
A typical basis of the Hilbert space
${\cal H}(K;N/D,M_{K})$ is written as
$\{c^{\dagger}_{Di_{1}+K}
\cdots
c^{\dagger}_{Di_{M_{K}}+K}|0>~|~
0\leq i_{1} < \cdots < i_{M_{K}} \leq \frac{N}{D}-1 \}$.
The operators $c^{\dagger}_{Di+K}$ and $c_{Dj+K}$ act only on the Hilbert
space ${\cal H}(K;N/D,M_{K})$ characterized by the
residue $K$ mod $D$.
We notice that the dimension of the Hilbert space is consistently equal to
\[
\sum_{M_{0}+\cdots+M_{D-1}=M}
(\prod_{K=0}^{D-1}{}_{N/D}C_{M_{K}})={}_{N}C_{M}.
\]

As far as the first part $\sum H_{K00}$ is only concerned, the
numbers $M_{0}$, $M_{1}$, $\cdots$, $M_{K}$ belong to good quantum
numbers so that they are determined simultaneously with energy
eigenvalues of $\sum H_{K00}$. At this point, we guess that the ground state
of $\sum H_{K00}$ is in the special Hilbert space
characterized by equal partition, that is,
$\bigotimes_{K=0}^{D-1}{\cal H}(K;\frac{N}{D},\frac{M}{D})$.
This conjecture can be justified by numerical calculations.
In fact, calculations with small number of electrons show that it is true.

When we apply the perturbation theory to our case,
it is necessary to study the structure of the Hilbert space ${\cal
H}(K;\frac{N}{D},\frac{M}{D})$. To do so,
we use symmetries of the Hamiltonian $H_{K00}$.

In order to make following discussions as simple as possible, we consider
only two operators $S_{K}$ and $T_{K}$,
defined as following, which commute with $H_{K00}$:
\begin{eqnarray*}
S_{K}c_{Dj+K}^{\dagger}&=& \exp(i\frac{2\pi}{qG}j)
c_{Dj+K}^{\dagger}S_{K}, ~~S_{K}|0>=|0>,\\
T_{K}c_{Dj+K}^{\dagger}&=& c_{D(j+1)+K}^{\dagger}T_{K}, ~~T_{K}|0>=|0>.
\end{eqnarray*}
Acting on the Hilbert space ${\cal H}(K;\frac{N}{D},\frac{M}{D})$,
the elementary operators $S_{K}$ and $T_{K}$ do not commute with each other, but obtain
the phase such as
\[
S_{K}T_{K}=\exp(i 2 \pi \frac{p}{q})T_{K}S_{K}.
\]

The Hilbert space ${\cal H}(K;\frac{N}{D},\frac{M}{D})$
is decomposed by using the symmetry group: $G_{\mbox{core}}
=\{ \xi^{l}T^{a}_{K}S^{b}_{K}~|
~l=0,\cdots,q-1;~a,b=0,\cdots,qG-1 \}$,
where $\xi^{q}=T^{qG}_{K}=S^{qG}_{K}=1=T^{\dagger}_{K}T_{K}=S^{\dagger}_{K}S_{K}$
and $S_{K}T_{K}=\xi T_{K}S_{K}$.
Finding representations of the group $G_{\mbox{core}}$, we obtain that
\[
{\cal H}(K;qG,pG)=
\bigoplus_{t,s=0}^{G-1}({\cal H}_{pts}(K))^{\oplus d_{ts}},
\]
where each Hilbert space ${\cal H}_{pts}(K)$ is $q$-dimensional.
Here, the point is that, without detailed information about the states in
${\cal H}_{pts}(K)$, it is possible to construct a low energy
effective Hamiltonian.

Actually, with the original couplings of $W(a,b)$, additional
symmetries appear so that degeneracy will be given by a multiple
of $q$. However, numerical studies show that the true ground state
in ${\cal H}(K;\frac{N}{D},\frac{M}{D})$ has the $q$-fold
degeneracy. This means that the ground state is still in one of ${\cal
H}_{pts}(K)$'s in the presence of additional symmetries.
Hence, it is not needed to consider representations of additional
symmetries as far as the ground state is concerned.
Further discussions on additional symmetries were given in Refs. 3 and 5.

A basis of the $q$-dimensional Hilbert space ${\cal H}_{pts}(K)$ is
given by $\{|K;t,s>,T_{K}|K;t,s>,\cdots,T^{q-1}_{K}|K;t,s>\}$.
The reference state $|K;t,s>$ plays a role like a highest weight state here.
The operators $T^{a}_{K}$ and $S^{b}_{K}$ act on $|K;t,s>$ such that
\begin{eqnarray*}
T_{K}^{mq+x}|K;t,s>&=&\exp(i\frac{2\pi}{G}mt)T_{K}^{x}|K;t,s>,\\
S_{K}^{nq+y}|K;t,s>&=&\exp(i\frac{2\pi}{G}ns)|K;t,s>,
\end{eqnarray*}
where $m$ and $n$ are integers, and $0\leq ~x,y~\leq q-1$.

Since there is no difference between $H_{K00}$'s except for the dummy index $K$,
the ground state of each Hamiltonian $H_{K00}$ lives in the
subspace, which is characterized by the same special values
of $t$ and $s$, regardless of $K$.
As a result, the ground state of the first part of the Hamiltonian, $\sum
H_{K00}$, lives in $\bigotimes_{K=0}^{D-1}{\cal H}_{p t s}(K)$.
We notice that the ground state of $\sum H_{K00}$ is degenerate as
many as $q^{D}$.
A basis of the $q^{D}$-dimensional Hilbert space
$\bigotimes_{K=0}^{D-1}{\cal H}_{pts}(K)$
is given by $\{ T_{0}^{i_{0}}T_{1}^{i_{1}}\cdots T_{D-1}^{i_{D-1}}|t,s>~|
~0\leq ~i_{0},\cdots,i_{D-1}~ \leq q-1 \}$,
where $|t,s>\equiv |0;t,s>\otimes\cdots\otimes |D-1;t,s>$.
In order to find the role of the other parts
for the ground state, we now apply the well-known perturbation theory to our $q^{D}$
degenerate case. Up to now, in this
analysis, we have reduced the dimension of
the relevant Hilbert space from ${}_{qGD}C_{pGD}$ to $q^{D}$.


The first thing to compute in the perturbation theory
for a degenerate case is the expectation value:
$<\psi_{i}|2\sum_{KA} H_{KA0}+\sum_{KAB}
H_{KAB}|\psi_{j}>$, where $|\psi_{i}>\in
\bigotimes_{K=0}^{D-1}{\cal H}_{pts}(K)$.
We notice that $<\psi_{i}|H_{KAB}|\psi_{j}>=0$
for any $i$ and $j$ because the number of electrons in each sector
is not conserved for this Hamiltonian operator.
Hence, up to the first order, the property of
$<\psi_{i}|H_{KA0}|\psi_{j}>$ is of main concern.

In order to find the expectation value, we manipulate the
Hamiltonian $H_{KA0}$.
Using the Fourier transformation such as
\[
W(a,b)=\sum_{\alpha=0}^{N-1}\tilde{W}(\alpha,b)\exp(i\frac{2\pi}{N}a\alpha),
\]
and the property of dummy index $a$,
we rewrite the Hamiltonian as
\begin{eqnarray*}
&&H_{KA0}
=-\sum_{b=0}^{qG-1}
\sum_{\alpha=0}^{N-1}\tilde{W}(\alpha,Db)
\exp(i\frac{2\pi}{N}A\alpha) \\
&&\times
\sum_{a=0}^{qG-1}\exp(i\frac{2\pi}{N}Da\alpha)
c_{Da+K+A}^{\dagger}
c_{D(a+b)+K+A} \\
&&\times
\sum_{k=0}^{qG-1}\exp(-i\frac{2\pi}{N}Dk\alpha)
c_{D(k+b)+K}^{\dagger}
c_{Dk+K}.
\end{eqnarray*}
The minus sign in front comes from the anti-commutation
relation between $c^{\dagger}$ and $c$.
Here, we introduce a short hand notation for the important operator:
\[
\sum_{k=0}^{qG-1}
\exp(-i\frac{2\pi}{N}Dk\alpha)
c_{D(k+b)+K}^{\dagger}
c_{Dk+K} \equiv
N_{\mbox{op}}(K;\alpha,b).
\]
It is important to notice
that the expectation value of $<\psi_{i}|H_{KA0}|\psi_{j}>$
is written in terms of the elementary values of
$<K;t,s|T_{K}^{-i_{K}}N_{\mbox{op}}(K;\alpha,b)T_{K}^{j_{K}}|K;t,s>$
where $T_{K}^{j_{K}}|K;t,s>\in {\cal H}_{pts}(K)$.
Thus, we focus our attention on the property of $N_{\mbox{op}}(K;\alpha,b)$.

We find the following commutation relations:
\begin{eqnarray*}
T_{K}N_{\mbox{op}}(K;\alpha,b)
&=&\exp(i\frac{2\pi}{N}D\alpha)N_{\mbox{op}}(K;\alpha,b)T_{K},\\
S_{K}N_{\mbox{op}}(K;\alpha,b)
&=&\exp(i\frac{2\pi}{N}Db)
N_{\mbox{op}}(K;\alpha,b)S_{K}.
\end{eqnarray*}
Using these commutation relations, we notice that
\[
N_{\mbox{op}}(K;\alpha,b){\cal H}_{pts}(K) \subset  ({\cal
H}_{p~t+\alpha~s+b}(K))^{\oplus d_{t+\alpha~s+b}},
\]
where the indices $t+\alpha$ and $s+b$ should be
considered as the residues mod $G$.
Hence, the properties of decomposition give the important result that
$N_{op}(K;\alpha,b){\cal H}_{pts}(K) \perp {\cal
H}_{pts}(K)$ unless $\alpha=G(q\Omega+x)$ and $b=Gy$,
where $0 \leq \Omega \leq D-1$
and $0 \leq ~x,y~ \leq q-1$. Therefore, as far as the expectation values are
concerned, it is enough to consider only $N_{\mbox{op}}(K;G(q\Omega+x),G y)$
which is equal to $N_{\mbox{op}}(K;G x,G y)$.
Taking the inverse Fourier transformation
for $\tilde{W}(G(q \Omega+x),D G y)$, the couplings in the Hamiltonian are rewritten
in terms of the original $W(a,b)$.

We rewrite $N_{\mbox{op}}(K;G x,G y)$
in terms of $S_{K}$ and $T_{K}$:
\begin{eqnarray*}
N_{\mbox{op}}(K;G x,G y) \rightarrow
f(x,y)T_{K}^{\tilde{y}}S_{K}^{-\tilde{x}},~~~~~~~~~\\
f(x,y)\equiv
<K;t,s|T_{K}^{-\tilde{y}}N_{\mbox{op}}(K;Gx,Gy)|K;t,s>,
\end{eqnarray*}
where $x=\mbox{mod}(p\tilde{x},q)$, $y=\mbox{mod}(p\tilde{y},q)$,
and $0 \leq ~\tilde{x},\tilde{y}~ \leq q-1$.
We notice that both operators have the same commutation relations
with $S_{K}$ and $T_{K}$.
In order to satisfy
\[
N_{\mbox{op}}(K+D;\alpha,b)=
\exp(i\frac{2\pi}{N}D\alpha)N_{\mbox{op}}(K;\alpha,b),
\]
we require the following condition for consistency:
\[
S_{K+D}\equiv \exp(-i 2\pi \frac{p}{q})S_{K},~~~~T_{K+D}=T_{K}.
\]
The Hermitian conjugate of $N_{\mbox{op}}(K+A;G x,G y)$ is also
written in terms of $S_{K+A}$ and $T_{K+A}$.

Summing up, we present the effective Hamiltonian
for the FQHE at $\nu=p/q$:
\[
H_{\mbox{eff}}(\frac{p}{q})=\sum_{A=1}^{D-1}\sum_{x,y=0}^{q-1}J(A;x,y)
\sum_{K=0}^{D-1}S_{K+A}^{\tilde{x}}T_{K+A}^{-\tilde{y}}T_{K}^{\tilde{y}}S_{K}^{-\tilde{x}},
\]
where the couplings are written as
\begin{eqnarray*}
J(A;x,y)=-2\frac{|f(x,y)|^{2}}{qG}~~~~~~\\
\times
\sum_{b=0}^{qG-1}
W(A+Db,D G y)
\exp(-i\frac{2\pi}{q}b x).
\end{eqnarray*}
This effective Hamiltonian satisfies the requirement
of Hermitian.

Let us see the explicit expression of the Hamiltonian for the case
of $p/q=1/2$. For odd $G\equiv 2n+1$,
the operator $S_{K+A}^{\tilde{x}}T_{K+A}^{-\tilde{y}}T_{K}^{\tilde{y}}S_{K}^{-\tilde{x}}$
is equivalent to $S_{K+A}^{Gx}T_{K+A}^{-Gy}T_{K}^{Gy}S_{K}^{-Gx}$,
because the operator
$S_{K+A}^{2nx}T_{K+A}^{-2ny}T_{K}^{2ny}S_{K}^{-2nx}$
commutes freely and
plays no role in the Hilbert space $\bigotimes_{K=0}^{D-1}{\cal H}_{1ts}(K)$.
Since $S_{K}^{-G}=S_{K}^{G}$ and $T_{K}^{-G}=T_{K}^{G}$ for $q=2$, we find that
the effective Hamiltonian for $p/q=1/2$ is presented as
\begin{eqnarray*}
&&~~~~H_{\mbox{eff}}(\frac{1}{2})=\sum_{A=1}^{D-1}\sum_{K=0}^{D-1}[J(A;1,0)S_{K+A}^{G}S_{K}^{G} \\
&&+J(A;0,1)T_{K+A}^{G}T_{K}^{G}+J(A;1,1)S_{K+A}^{G}T_{K+A}^{G}T_{K}^{G}S_{K}^{G}].
\end{eqnarray*}
We notice that
the commutation relations between $L_{1}\equiv \frac{1}{2}S_{K}^{G}$,
$L_{2}\equiv \frac{1}{2}T_{K}^{G}$ and $L_{3}\equiv
\frac{i}{2}T_{K}^{G}S_{K}^{G}$,
are the same as those of SU(2),
$[L_{i},L_{j}]=i\epsilon_{ijk}L_{k}$.
Hence, we can call the Hamiltonian $H_{\mbox{eff}}(\frac{1}{2})$
with odd $G$ as a long range one-dimensional XYZ-model with
spin $\frac{1}{2}$.


Let us discuss symmetries of the effective Hamiltonian, in order to
decompose the Hilbert space.
We find several operators, which not only commute with the effective
Hamiltonian but also preserve the commutation relation of
$S_{K}T_{K}=\exp(i2\pi\frac{p}{q})T_{K}S_{K}$. As a result, we
notice the symmetries:
\begin{eqnarray*}
&&~~~~~~~~~~~~~~S\equiv \prod_{K=0}^{D-1}S_{K},~~~T\equiv \prod_{K=0}^{D-1}T_{K},\\
&&~~~~Q \left( \begin{array}{c}
         S_{K} \\
         T_{K}
        \end{array}   \right)
=\left( \begin{array}{c}
         S_{K+1} \\
         T_{K+1}
        \end{array}   \right) Q,~~~Q|t,s>=|t,s>.
\end{eqnarray*}
Since we notice that 
$|f(x,y)|^{2}=|f(-x,-y)|^{2}$ from the
relation
\[
f^{\ast}(x,y)=\exp(i\frac{2\pi}{q}x(\tilde{y}-Gy))f(-x,-y),
\]
there exists the additional symmetry,
\begin{eqnarray*}
P\left( \begin{array}{c}
         S_{K} \\
         T_{K}
        \end{array}   \right)
=\left( \begin{array}{c}
         S_{D-K}^{-1} \\
         T_{D-K}^{-1}
        \end{array}   \right) P,~~~P|t,s>=|t,s>.
\end{eqnarray*}

The commutation relations between these elementary operators are given by
\begin{eqnarray*}
&&ST=\exp(i 2\pi \frac{p}{q}D )TS,~~SQ=QS,~~TQ=QT,\\
&&SP=PS^{-1},~~TP=PT^{-1},~~QP=PQ^{-1}.
\end{eqnarray*}
Since the method of topological invariance suggests that the ground
state of the FQHE is $q$-degenerate, we require that $q$ and $D$
are coprime numbers: $\xi\equiv \exp(i2\pi pD/q)$, $\xi^{q}=1$. We now consider the
symmetry group:
$G_{q}=\{\xi^{l}P^{n}Q^{d}T^{a}S^{b}|l=0,\cdots,q-1;~n=0,1;
~d=0,\cdots,D-1;~a,b=0,\cdots,qG-1\}$.
Representations of this group
are essential in order to understand the
structure of spectrum.

We first consider the following characters 
for a group element $g=\xi^{l}P^{n}Q^{d}T^{a}S^{b}$:
\[
\Psi_{xkts}(g) =\left \{
   \begin{array}{ll}
     q_{x}\zeta_{q}^{xl}(\zeta_{D}^{kd}\zeta_{qG}^{ta+sb}+\mbox{c.c.})
              & \mbox{if}~n=0,~q_{x}|a,b \\
     0 & \mbox{otherwise,}
   \end{array}\right.
\]
where $x=0,\cdots,q-1$; $k=0,\cdots,D-1$; $t,s=0,\cdots,qG/q_{x}-1$;
$q_{x}=q/\mbox{gcd}(q,x)$, and $\zeta_{y}\equiv\exp(i 2\pi / y)$.
In order to avoid double counting, we should notice that
$\Psi_{xkts}=\Psi_{x~(D-k)~(qG/q_{x}-t)~(qG/q_{x}-s)}$.
Then, it is found that all $\Psi_{xkts}$ are irreducible characters
except for $\Psi_{x~(\alpha D/2)~(\beta qG/2q_{x})
~(\gamma qG/2q_{x})}$, where $\alpha,\beta,\gamma=0,1$.
Of course, the cases containing $D/2$ or $qG/2q_{x}$ are
meaningful only when they are integers. We find the irreducible
components,
\[
\Psi_{x~(\alpha D/2)~(\beta qG/2q_{x})
~(\gamma qG/2q_{x})}=\chi^{+}_{x\alpha\beta\gamma}+\chi^{-}_{x\alpha\beta\gamma},
\]
\[
\chi^{\pm}_{x\alpha\beta\gamma}(g) =\left \{
   \begin{array}{ll}
     q_{x}\zeta_{q}^{xl}(\pm 1)^{n}
      (-1)^{\alpha d +\beta\frac{a}{q_{x}}+\gamma \frac{b}{q_{x}}}
          & \mbox{if}~q_{x}|a,b \\
       0 & \mbox{otherwise.}
   \end{array}\right.
\]

The reducible character of the symmetry group for the Hilbert space
$\bigotimes_{K=0}^{D-1}{\cal H}_{p\tilde{t}\tilde{s}}(K)$ is found as
\[
\chi(\xi^{l}Q^{d}T^{a}S^{b}) =\left \{
   \begin{array}{ll}
     q^{\mbox{gcd}(D,d)}
     \zeta_{q}^{p^{\prime}l}\zeta_{qG}^{t^{\prime}a+s^{\prime}b}
            & \mbox{if}~q|a,b \\
     0 & \mbox{otherwise,}
   \end{array}\right.
\]
where $p^{\prime}=\mbox{mod}(Dp,q)$, $t^{\prime}=\mbox{mod}(D\tilde{t},G)$,
and $s^{\prime}=\mbox{mod}(D\tilde{s},G)$.
We omit presenting $\chi(\xi^{l}PQ^{d}T^{a}S^{b})$, which is not
necessary in the following discussion.

Using the irreducible characters $\Psi_{xkts}$
and the reducible character $\chi$, we decompose the Hilbert space
as
\[
\bigotimes_{K=0}^{D-1}{\cal H}_{p\tilde{t}\tilde{s}}(K)=
\bigoplus_{k,t,s}({\cal H}_{p^{\prime}kts}^{\oplus m_{kts}}).
\]
The multiplicity $m_{kts}$ is calculated as
\begin{eqnarray*}
&&m_{kts}=(\Psi_{p^{\prime}kts},\chi)\\
&&=\frac{1}{2qD}\sum_{d=0}^{D-1}q^{\mbox{gcd}(D,d)}
(\zeta_{D}^{-kd}\delta_{t,t^{\prime}}\delta_{s,s^{\prime}}+
\zeta_{D}^{kd}\delta_{t,G-t^{\prime}}\delta_{s,G-s^{\prime}}).
\end{eqnarray*}

Numerical calculations show that the ground state for $H_{K00}$ exists
usually in the sector of $(\tilde{t},\tilde{s})=(0,0)$ so
$(t^{\prime},s^{\prime})=(0,0)$.
Hence, non-zero values of $m_{kts}$ are given for $(t,s)=(0,0)$.
For instance, we notice the consistency for dimensionality:
$2q\times \sum_{k=1}^{D/2-1}m_{k00}+q\times m_{000}+q\times m_{\frac{D}{2}00}=q^{D}$.
This decomposition will be useful in numerical calculations,
especially, the Lanczos method.


In summary, we have derived the effective Hamiltonian from the
Yoshioka-Halperin-Lee Hamiltonian.
It is remarkable that the effective Hamiltonian
is written as the similar form as that of the Heisenberg chain.
For this reason, we can call the model described by the effective
Hamiltonian as ``quantum Hall chain".
The Haldane gap in the Heisenberg chain has been
studied by various numerical methods:
exact diagonalizations\cite{6,7}, quantum Monte Carlo\cite{8,9}, and
density matrix renormalization group\cite{10}.
We suggest that similar numerical hard works are needed to analyze our
system of quantum Hall chain.
The numerical works should confirm that the gap appears for odd $q$, while
no gap for even $q$\cite{11}.
Conclusive reports on the gap of quantum Hall chain are anticipated.

\section*{Acknowledgment}

This work was supported in part by the Basic Science Research Institute
Program, Ministry of Education, Project No. 98-015-D00061.
The authors are grateful to Jin Hong for helpful discussions.
M.H.C. wishes to thank Professor Dr.
Reinhard Oehme for his encouragement.

\end{multicols}

\end{document}